\documentclass[%
 aip,
 amsmath,amssymb,superscriptaddress,
 reprint%
]{revtex4-1}

\usepackage{graphicx}
\usepackage{caption}
\usepackage{xcolor}
\usepackage{dcolumn}
\usepackage{bm}
\usepackage{subfig}

\usepackage[utf8]{inputenc}
\usepackage[T1]{fontenc}
\usepackage{mathptmx}
\usepackage{etoolbox}
\usepackage{comment}

\makeatletter
\def\@email#1#2{%
 \endgroup
 \patchcmd{\titleblock@produce}
  {\frontmatter@RRAPformat}
  {\frontmatter@RRAPformat{\produce@RRAP{*#1\href{mailto:#2}{#2}}}\frontmatter@RRAPformat}
  {}{}
}%
\makeatother
\begin{document}

\preprint{AIP/123-QED}

\title[Phase dynamics of tunnel Al-based ferromagnetic Josephson junctions]{Phase dynamics of tunnel Al-based ferromagnetic Josephson junctions}
\author{H.G. Ahmad}
\email{halimagiovanna.ahmad@unina.it}
\affiliation{ 
Dipartimento di Fisica “Ettore Pancini”, Università degli Studi di Napoli Federico II, Via Cinthia, I-80126, Napoli, Italy
}
\author{R. Satariano}
\affiliation{ 
Dipartimento di Fisica “Ettore Pancini”, Università degli Studi di Napoli Federico II, Via Cinthia, I-80126, Napoli, Italy
}
\affiliation{CNR-SPIN, UOS Napoli, Monte S. Angelo, via Cinthia, I-80126, Napoli, Italy}

\author{R. Ferraiuolo}
\affiliation{ 
Dipartimento di Fisica “Ettore Pancini”, Università degli Studi di Napoli Federico II, Via Cinthia, I-80126, Napoli, Italy
}
\affiliation{CNR-SPIN, UOS Napoli, Monte S. Angelo, via Cinthia, I-80126, Napoli, Italy}

\author{A. Vettoliere}
\affiliation{ 
Consiglio Nazionale delle Ricerche-ISASI, Via Campi Flegrei 34, I-80078 Pozzuoli, Italy
}
\author{C. Granata}
\affiliation{ 
Consiglio Nazionale delle Ricerche-ISASI, Via Campi Flegrei 34, I-80078 Pozzuoli, Italy
}
\author{D. Montemurro}
\affiliation{ 
Dipartimento di Fisica “Ettore Pancini”, Università degli Studi di Napoli Federico II, Via Cinthia, I-80126, Napoli, Italy
}
\author{G. Ausanio}
\affiliation{ 
Dipartimento di Fisica “Ettore Pancini”, Università degli Studi di Napoli Federico II, Via Cinthia, I-80126, Napoli, Italy
}
\affiliation{CNR-SPIN, UOS Napoli, Monte S. Angelo, via Cinthia, I-80126, Napoli, Italy}

\author{L. Parlato}
\affiliation{ 
Dipartimento di Fisica “Ettore Pancini”, Università degli Studi di Napoli Federico II, Via Cinthia, I-80126, Napoli, Italy
}
\affiliation{CNR-SPIN, UOS Napoli, Monte S. Angelo, via Cinthia, I-80126, Napoli, Italy}
\author{G.P. Pepe}
\affiliation{ 
Dipartimento di Fisica “Ettore Pancini”, Università degli Studi di Napoli Federico II, Via Cinthia, I-80126, Napoli, Italy
}
\affiliation{CNR-SPIN, UOS Napoli, Monte S. Angelo, via Cinthia, I-80126, Napoli, Italy}

\author{F. Tafuri}
\affiliation{ 
Dipartimento di Fisica “Ettore Pancini”, Università degli Studi di Napoli Federico II, Via Cinthia, I-80126, Napoli, Italy
}

\author{D. Massarotti}
\affiliation{ 
Dipartimento di Ingegneria Elettrica e delle Tecnologie dell’Informazione, Università degli Studi di Napoli Federico II, I-80125 Napoli, Italy
}
\affiliation{CNR-SPIN, UOS Napoli, Monte S. Angelo, via Cinthia, I-80126, Napoli, Italy}

\date{\today}

\begin{abstract}

By measuring the current-voltage characteristics and the switching current distributions as a function of temperature, we have investigated the phase dynamics of Al tunnel ferromagnetic Josephson junctions (JJs), designed to fall in the typical range of parameters of state-of-the-art transmons, providing evidence of phase diffusion processes. The comparison with the experimental outcomes on non-magnetic JJs with nominally the same electrodynamical parameters demonstrates that the introduction of ferromagnetic barriers does not cause any sizeable detrimental effect, and supports the notion of including tunnel ferromagnetic JJs in qubit architectures. 

\end{abstract}

\maketitle

The competition between the superconducting and the ferromagnetic order parameters in ferromagnetic Josephson junctions (JJs) establishes a wide range of interesting phenomena that can be exploited in superconducting electronics~\cite{ryazanov2001,Golubov2004,Buzdin2005,Bergeret2005,Weides2006,Eschrig2008,Robinson2010,Sickinger2012,baek2014,Singh2015,Linder2015,Gingrich2016,Dahir2019,Feofanov2010,Larkin2012,satchell2020,Satariano2021,Ahmad2022}. SIsFS (Superconductor-Insulator-superconducting interlayer-Ferromagnet-Superconductor) JJs, comprising both an insulating (I) and a ferromagnetic (F) layer, are typically characterized by an underdamped or moderately damped behavior~\cite{Weides2006,Larkin2012,Parlato2020,Karelina2021}, and may offer key advantages in some applications compared to standard metallic ferromagnetic JJs~\cite{Ahmad2022ferro}. Specifically, we proposed integrating tunnel ferromagnetic JJs in transmon qubit architectures~\cite{Ahmad2022ferro}. The \emph{ferrotransmon} has been proposed as a scalable alternative to flux-tunable transmon qubits (split-transmons) that use flux-tunable DC-SQUIDs~\cite{Krantz2019}, which have the disadvantage of being intrinsically bulky and sensitive to flux-noise fluctuations~\cite{clarke2006squid,granata2016,Krantz2019,Ahmad2022ferro,Ahmad2023}.

It is common opinion that superconducting qubit research began in the 1980s motivated by the question of whether macroscopic variables would behave in a quantum mechanical fashion~\cite{leggett1984, Devoret1985,Martinis1985PRL, Martinis1987,Martinis2009}. Initial experiments have verified the quantum behavior of JJs through the tunneling out of the zero-voltage state (Macroscopic Quantum Tunneling (MQT)) of a current-biased JJ and by the observation of Energy Level Quantization (EQL)~\cite{leggett1984, Devoret1985, Martinis1985PRL, Martinis1987, Clarke1988,Martinis2009}. These have inspired a series of later studies on JJs composed of a variety of materials and in quite different conditions for the junction electrodynamics parameters~\cite{Vion1996,kivioja2005,Bauch2005,mannik2005, Krasnov2007,longobardi2011,longobardi2012, Stornaiuolo2013, massarotti2015macro,massarotti2018electrodynamics,manucharyan2012evidence,haxell2023measurements} and have promoted the first phase qubit~\cite{Martinis2002,Martinis2009}. For an extensive review of most experiments, we refer to~\cite{Massarotti2019}. 

A full account of the phase dynamics of a JJ is the first step for whatever applicative direction and specifically for superconducting qubit architectures~\cite{Krantz2019,gumucs2023calorimetry, Lu2023}. The phase dynamics of JJs is essential to understand the impact of dissipation and noise fluctuations on the coherence performances of superconducting quantum devices and to distinguish contributions to dissipation due to the environment or the junction itself~\cite{Caldeira1981,Martinis1985PRL,Martinis1987,Kautz1990,Vion1996,Massarotti2019}. Switching current distribution (SCD) measurements are a unique tool for a complete account of phase dynamics of JJs\cite{Fulton1974} and have been used to demonstrate MQT and ELQ~\cite{Devoret1985, Martinis1987, Clarke1988}. It is well-known that the switching from the superconducting to the resistive states in a JJ is a stochastic process~\cite{Ivanchenko1968, Fulton1974, Caldeira1981,buttiker1983,leggett1984, Devoret1985, Martinis1987, Clarke1988, Garg1995, Vion1996,Massarotti2019}. Hence, by repeatedly ramping the JJ with a well-defined current-bias ramp, the switching currents will distribute according to the effects of thermal or quantum fluctuations. The mean value of the SCD, i.\;e., the average switching current $I_{sw}$, the SCD standard deviation $\sigma$, which weights the level of noise fluctuations occurring in the system, and the SCD skewness $\gamma$, which weights the asymmetry of the SCD, follow well-defined behaviors~\cite{Fulton1974, Devoret1985, Martinis1987, Clarke1988,fenton2008}. Thermal noise fluctuations, for example, increase the SCD standard deviation when increasing the temperature, while they saturate below a crossover temperature $T_{cr}$,
\begin{equation}
\label{Tcr}
    T_{cr}=\frac{\hbar\omega_P}{2\pi k_B}\left(\sqrt{1+\left(\frac{1}{2Q}\right)^2}-\frac{1}{2Q}\right),
\end{equation}
where $Q$ is the JJ quality factor
\begin{equation}
\label{Q}
Q=\omega_PCR,
\end{equation}
and $\omega_P$ is the JJ plasma frequency in the presence of a bias current $I$:
\begin{equation}
\label{plasma}
\omega_P=\left(1-(I/I_{c0})^2\right)^{1/4}\sqrt{\frac{2eI_{c0}}{\hbar C}}.
\end{equation}
Here $I_{c0}$ represents the critical current in the absence of fluctuations and $C$ is the capacitance of the JJ. $T_{cr}$ signals the transition from the Thermal Activation (TA) regime to the MQT one~\cite{leggett1984, Devoret1985, Martinis1985PRL, Martinis1987, Clarke1988}. In both cases, the SCD skewness is typically negative, and close to $-1$, indicating the presence of a characteristic tail for current-bias $I<I_{sw}$. In the moderately damped regime, for intermediate levels of dissipation, phase diffusion (PD) processes occur and distinctive fingerprints arise~\cite{Iansiti1987, Kautz1990, Vion1996, kivioja2005, mannik2005, krasnov2005,fenton2008, longobardi2011}: i)  $\sigma$ tends to decrease while increasing the temperature and SCDs become narrower; ii) $\gamma$ reaches values of the order of $0$, hence SCDs tend to symmetrize. 

In the present work, we have performed SCD measurements on Al-based tunnel ferromagnetic SIsFS JJs with electrodynamical parameters compatible with transmon qubits~\cite{Ahmad2022ferro,vettoliere2022, Vettoliere2022aluminum}, providing a fair characterization of phase dynamics of ferromagnetic tunnel JJs. The analyzed SIsFS JJs represent a milestone towards the implementation of the ferrotransmon, being the first tunnel ferromagnetic JJs compatible with standard fabrication technologies for transmon qubits, which are typically based on Al/Al\text{O${}_x$}/Al non-magnetic tunnel JJs characterized by critical currents of the order of few nanoamperes~\cite{Majer2007,DiCarlo2009,Oliver2013,Krantz2019}. The transport properties of SIsFS JJs are governed by a series between the tunnel SIs and the ferromagnetic sFS JJs, where F is a thin permalloy (Py) ferromagnetic barrier~\cite{vettoliere2022,Vettoliere2022aluminum}. The former guarantees the low-dissipation and high-quality factors required in highly coherent devices~\cite{Ahmad2020}, while the latter allows for magnetic switching and novel tunability schemes of the Josephson energy. 

By comparing the experimental data for the SIsFS JJs with non-magnetic SIS JJs built with nominally the same electrodynamical parameters, we demonstrate that the addition of the ferromagnetic layer does not influence the dissipation mechanisms of the devices. Both magnetic and non-magnetic JJs have shown PD effects, which are well-known to occur also in low-$E_J$ conventional non-magnetic JJs typically used in superconducting transmon qubits~\cite{Lu2023}. Additionally, we also report on the appearance of a finite resistance $R_0$ in the superconducting branch of the I-V characteristics for temperature higher than $T_c/2$, where $T_c$ is of the order of $1$ K, which is another fingerprint of PD processes in JJs characterized by low values of $E_J$~\cite{Iansiti1987,Kautz1990}. 
\begin{table}
\caption{\label{tab1}Summary for the Josephson energy $E_J$, the critical current $I_c$, the normal resistance $R_N$ and $\Gamma$ determined at base temperature of $10$ mK for SIS and SIsFS JJs of batches A and B.}\begin{ruledtabular}
\begin{tabular}{ccccc}
 & Batch A & & Batch B &\\
\hline
& SIS & SIsFS &  SIS & SIsFS \\
$E_J$ (K)& $0.86\pm0.01$ & $0.48\pm0.01$ & $2.28\pm0.02$ &$1.56\pm0.02$ \\
$I_c$ (nA)& $37.6\pm0.4$ & $21.6\pm0.2$ & $106\pm1$ & $65.8\pm0.7$ \\
$R_N$ (\text{k$\Omega$}) & $1.69\pm0.07$ & $3.41\pm0.14$ & $0.68\pm0.03$ & $1.41\pm0.06$ \\
$\Gamma$& $0.012$ & $0.015$ & $0.003$ & $0.004$ \\
\end{tabular}
\end{ruledtabular}
\end{table}
\begin{figure}
\subfloat[][\centering]{\includegraphics[width=0.85\columnwidth]{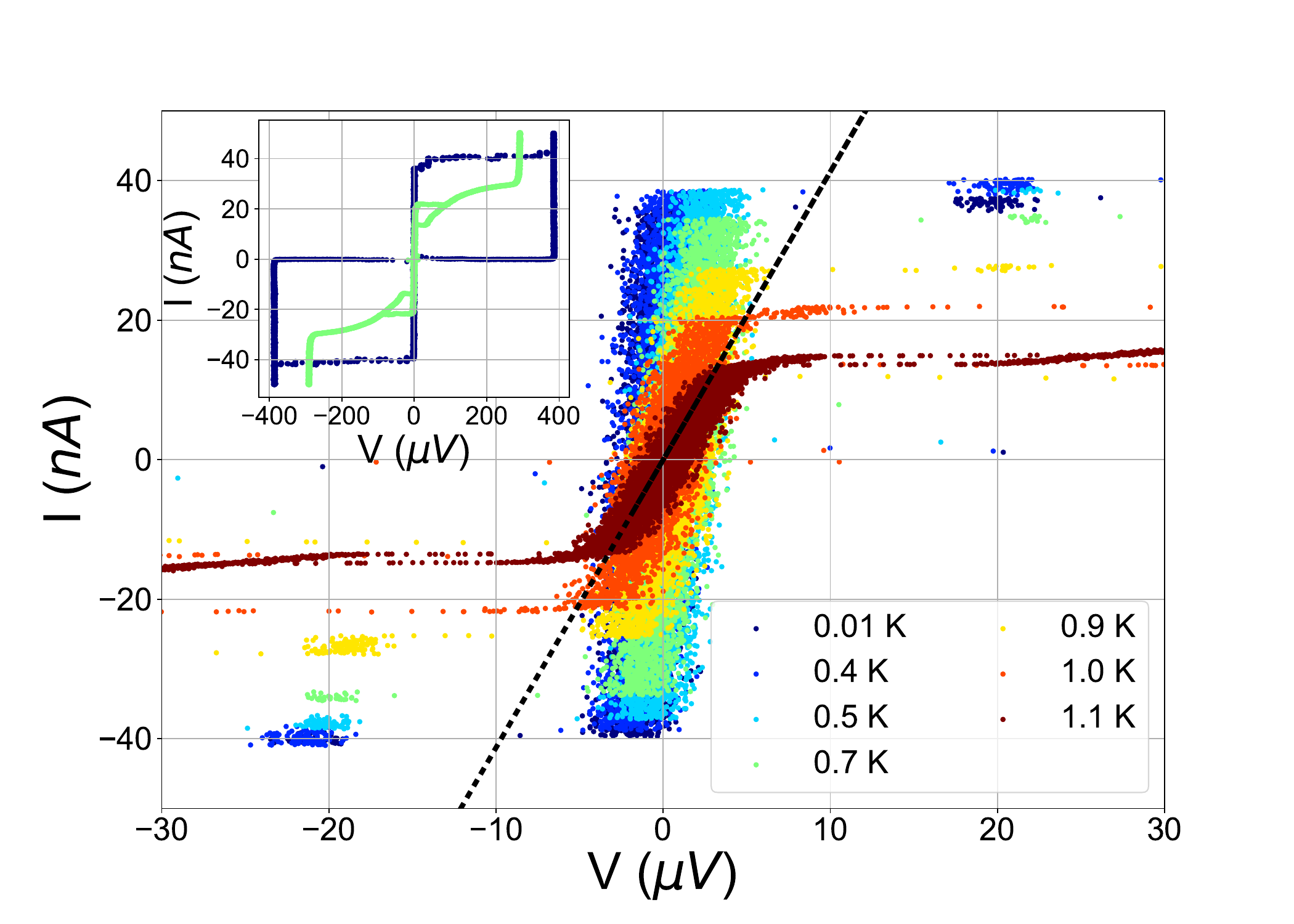}}\\
\subfloat[][\centering]{\includegraphics[width=0.85\columnwidth]{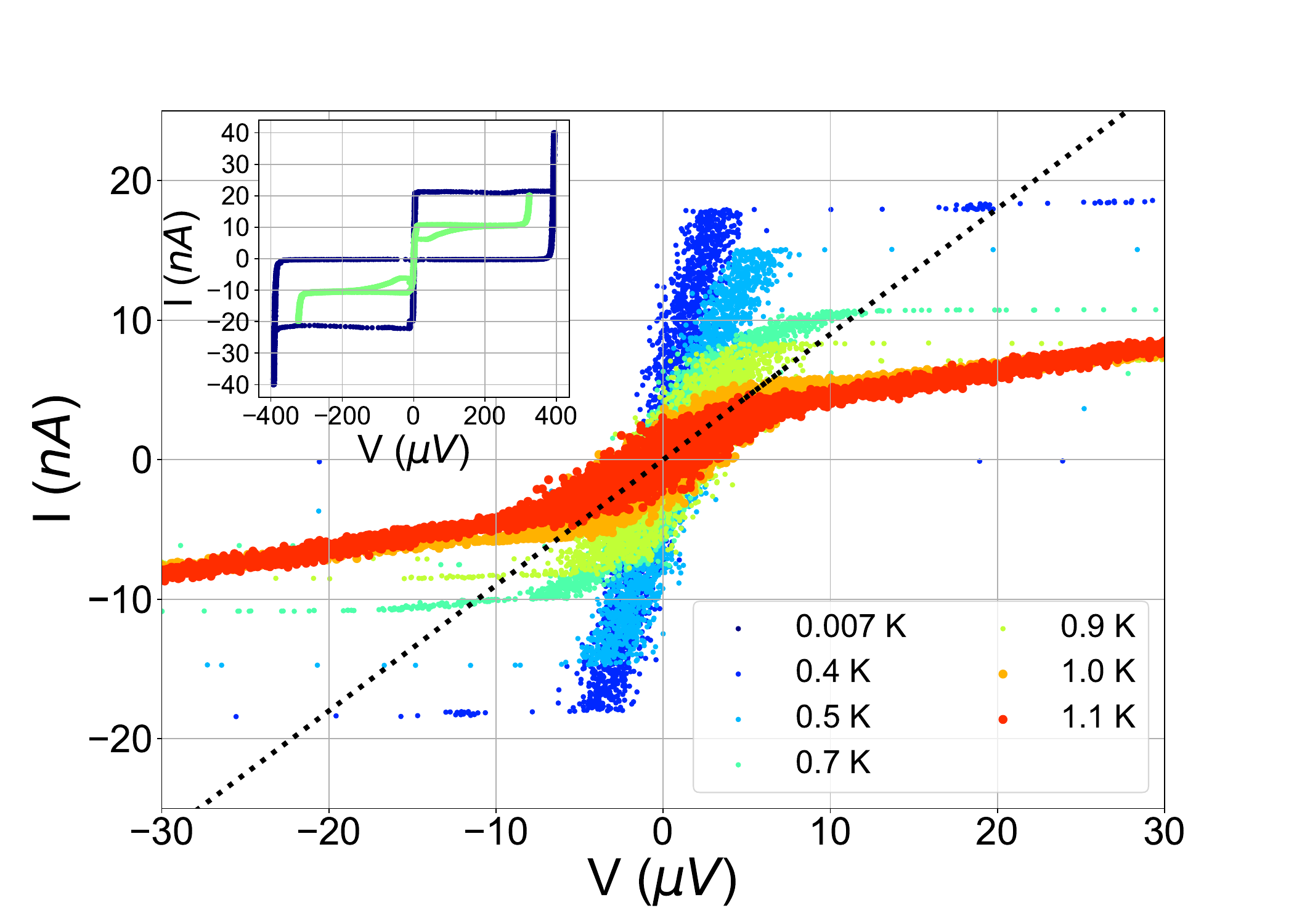}}\\
\subfloat[][\centering]{\includegraphics[width=0.85\columnwidth]{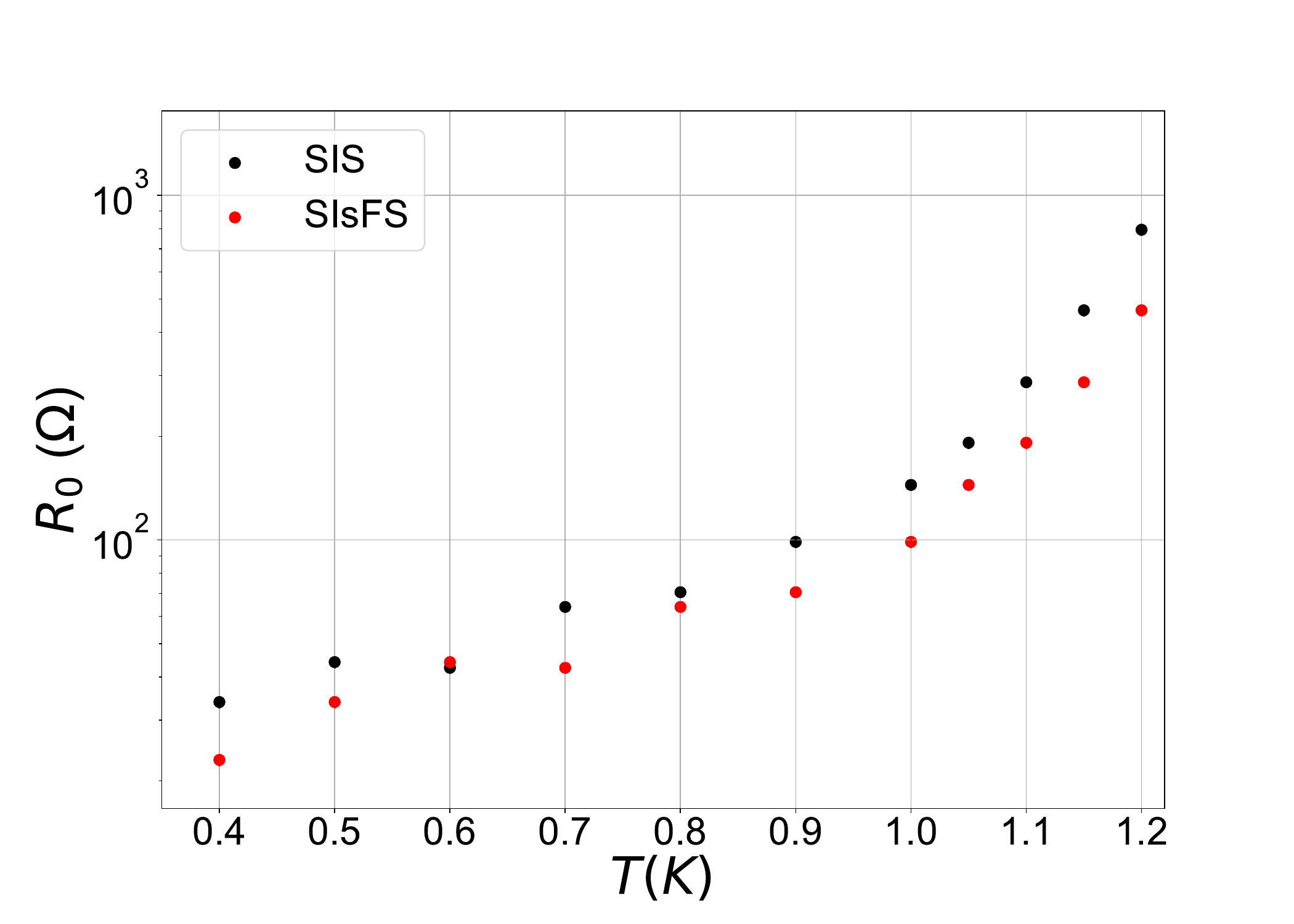}}
\caption{Evidence of a resistive branch in the I-V characteristics vs. temperature for Batch A. In panels a) and b), I-V superconducting branches for the SIS and the SIsFS JJs, respectively, where the dashed black line refers to the I-V phase-diffusion slope. In the insets, the I-V curves at the base and high temperatures are shown. The relative errors on measured currents and voltages are 1\% and 2\%, respectively. In panel c), finite slope of the supercurrent branch of the I-V curves $R_0$ vs. $T$ in the high-temperature range for the SIS (in black) and SIsFS JJs (in red) is shown, where $R_0$ has been estimated through the linear fitting of the I-V superconducting branch in panels a) and b). The relative error on $R_0$ is $4\%$.}
\label{fig1}
\end{figure}
\begin{figure*}[t]
\subfloat[][\centering]{\includegraphics[width=0.35\textwidth]{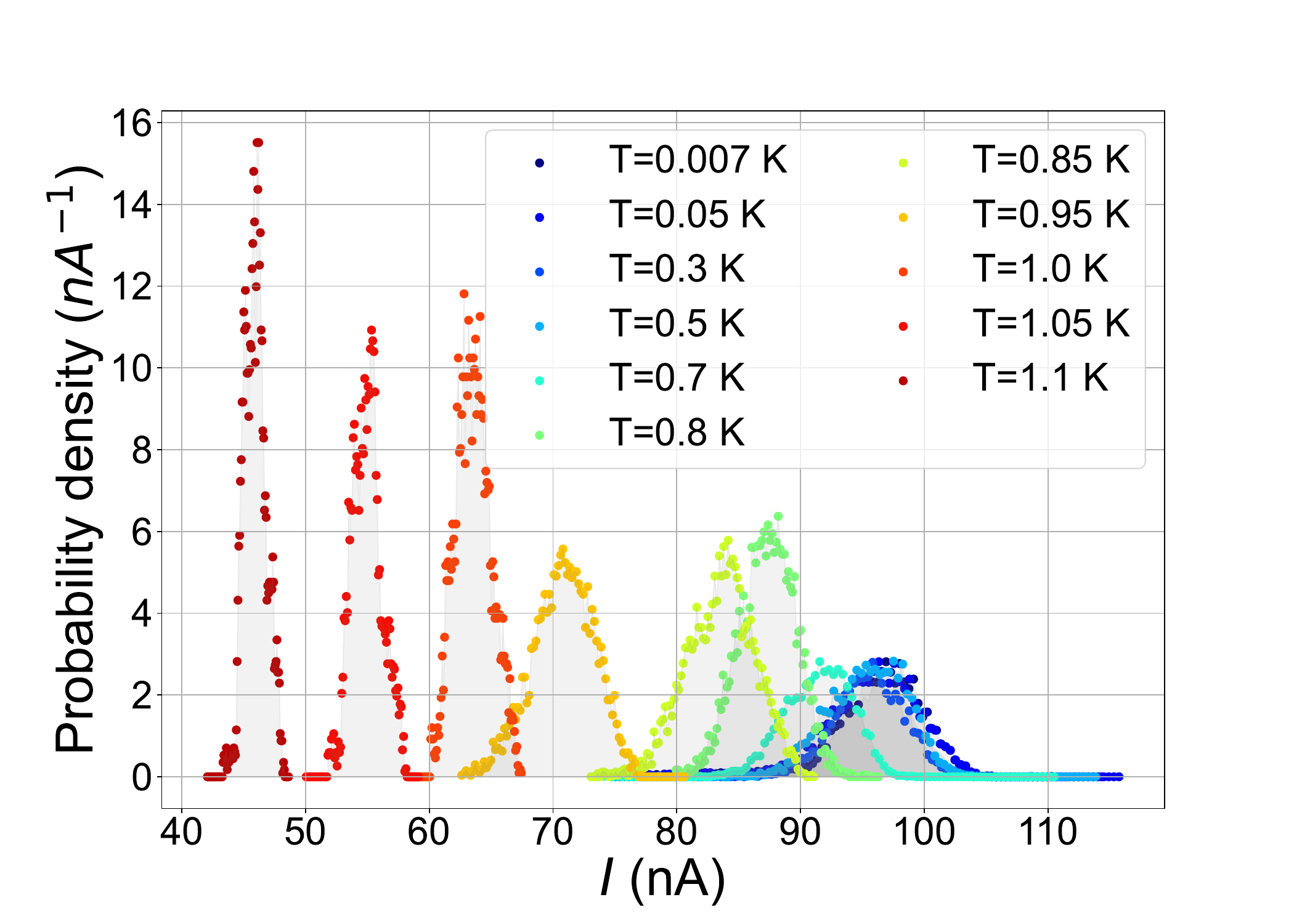}}
\subfloat[][\centering]{\includegraphics[width=0.35\textwidth]{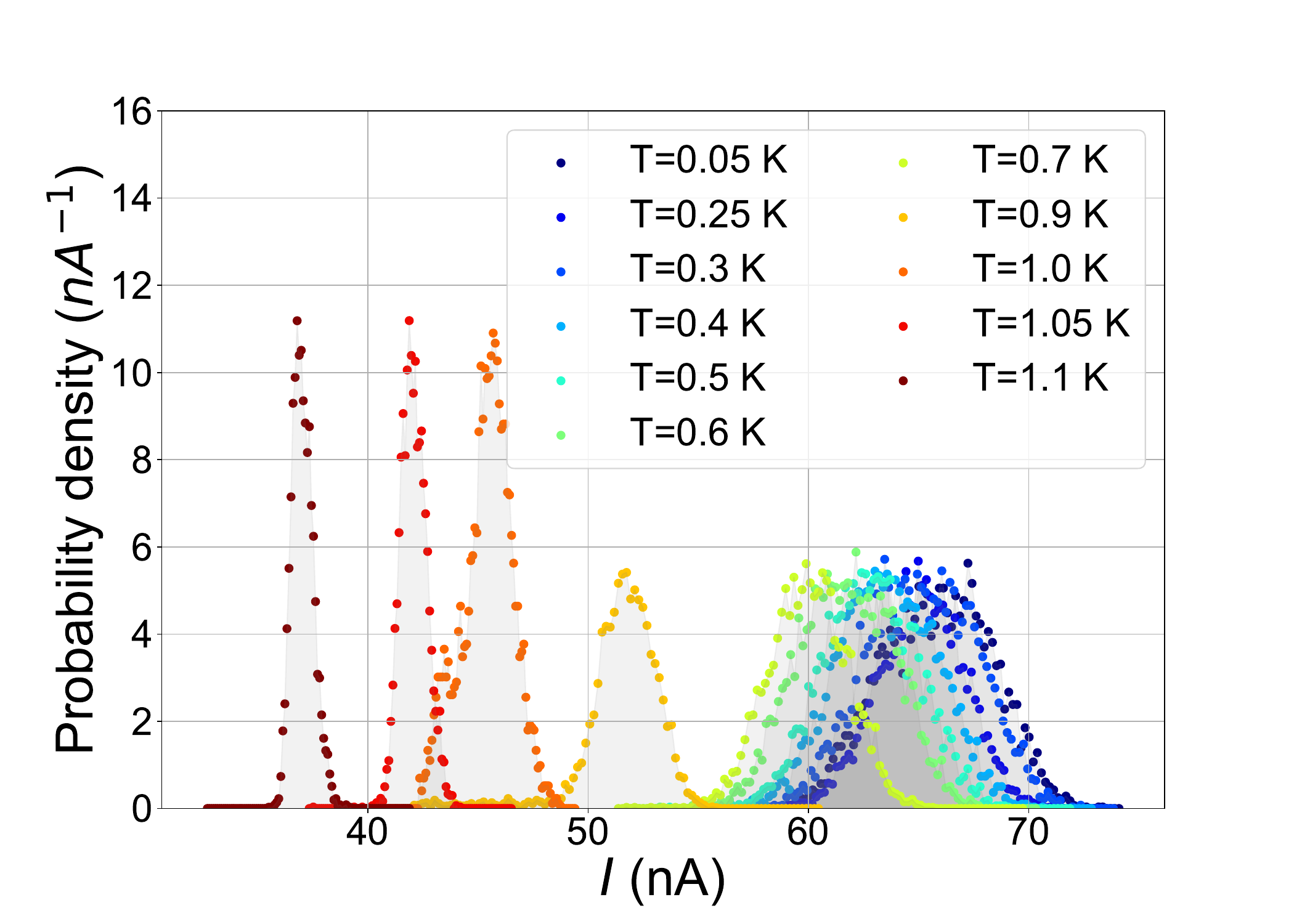}}
\subfloat[][\centering]{\includegraphics[width=0.35\textwidth]{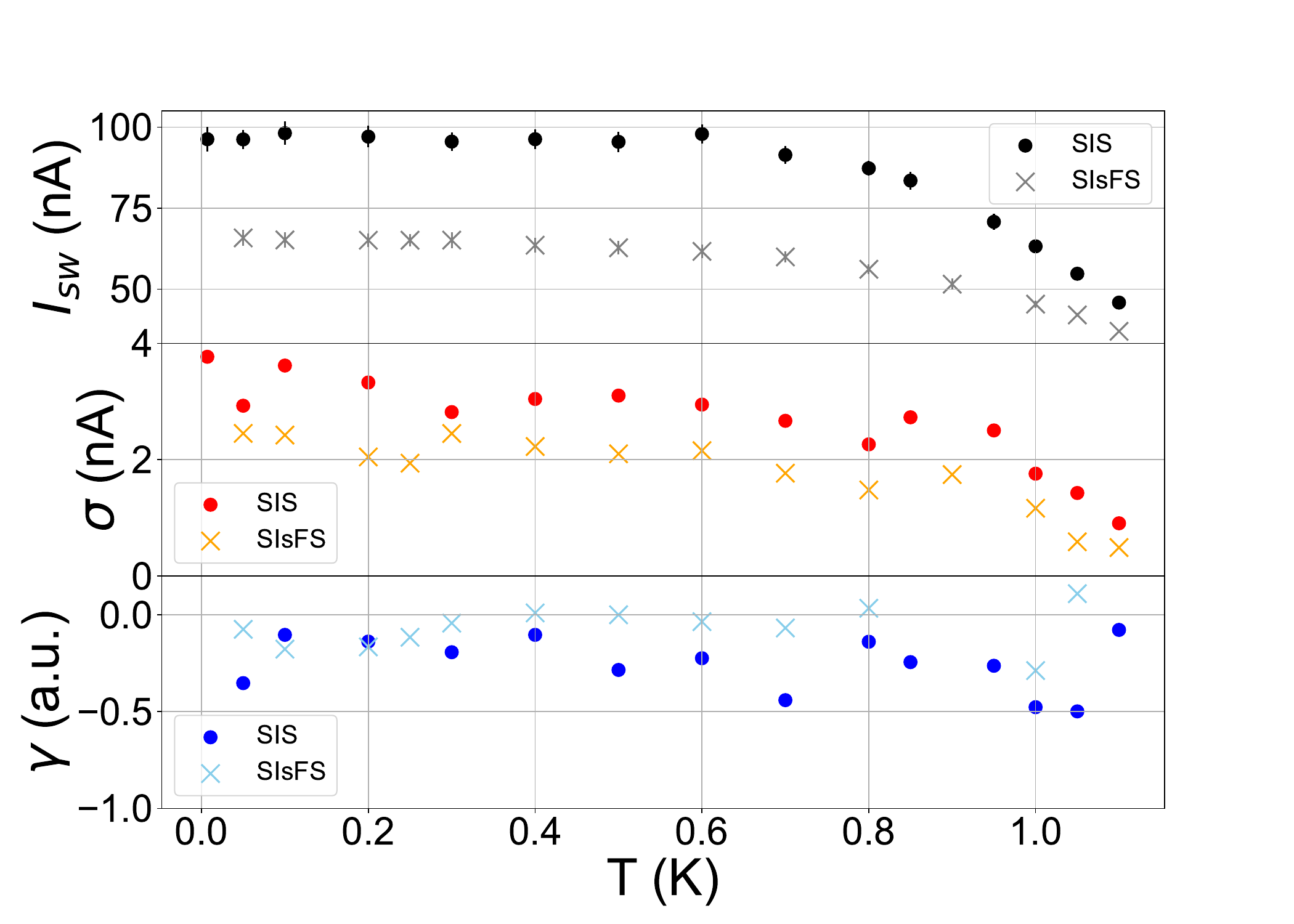}}
\caption{Switching Current Distributions as a function of temperature for the non-magnetic SIS JJ and the magnetic SIsFS JJ in batch B. In panels a) and b), we report the switching current probability densities for the two JJs, respectively. The mean value of $I_{sw}$, the standard deviation $\sigma$, and the skewness $\gamma$ of the SCDs as a function of temperature $T$ are reported in panel c). Circle and cross points refer to the SIS and the SIsFS JJs, respectively. }
\label{fig2}
\end{figure*}

We have measured two sample batches, namely batches A and B, respectively, including one circular tunnel SIsFS ferromagnetic JJ and one circular SIS non-magnetic tunnel JJ, taken as a reference. The nominal diameter of the JJs is $4$ \text{$\mu$m}, and thicknesses of the top and bottom electrodes are $350$ nm and $200$ nm, respectively. The thicknesses of the superconducting interlayers in the SIsFS JJs is $30$ nm, while the Py ($83\pm3\%$ Ni and $17\pm3\%$ Fe) and Al\text{$O_x$} barriers are of $3$ nm and $\sim1$ nm, respectively~\cite{vettoliere2022,Vettoliere2022aluminum}. The JJs have been anchored to the coldest stage of a dry dilution refrigerator with a nominal base temperature of $10$ mK (see Sec. 1 in Supplementary Material). To meet future use in ferrotransmon, junctions are designed and built to be in the $E_J$ regime of conventional transmon devices, summarized for convenience in Tab.~\ref{tab1}, which also includes the ratio $\Gamma=k_BT/E_J$ at base temperature. The nominal charging energy can be derived by considering the geometrical parallel-plate capacitance of the barrier, which is of the order of $C=350$ fF~\cite{vettoliere2022,Vettoliere2022aluminum}. This is well above typical capacitance values of conventional transmon devices, given that JJs here analyzed have much larger areas~\cite{koch2007}. 

In the PD regime, continuous escape and re-trapping of the phase-particle along the washboard potential follows a Brownian motion, which can be modeled through a Monte Carlo approach introducing a temperature-dependent random current-noise source in the Resistively and Capacitively-Shunted Junction (RCSJ) model~\cite{longobardi2011,longobardi2012,Stornaiuolo2013,Massarotti2015}. The inclusion of this stochastic noise leads to the phase-diagram $(Q,\Gamma)$ reported in Ref.~\cite{longobardi2012}, which allows to point out a direct correlation between the thermal energy $k_BT$, the Josephson potential energy $E_J$ and the sensitivity to noise and dissipation of a JJ, quantified through its quality factor $Q$ in Eq.~\ref{Q}~\cite{longobardi2012}. 

As discussed in Refs.~\cite{Iansiti1987,Kautz1990,Vion1996,Stornaiuolo2013}, in addition to the decrease of $\sigma$ when increasing the temperature, another experimental benchmark that can demonstrate PD is the presence of a finite resistance in the superconducting branch of the I-V characteristics. In Fig.~\ref{fig1} a) and b), we report evidence of a finite resistance $R_0$ in the superconducting branches of the I-V curves measured as a function of temperature for both non-magnetic and ferromagnetic tunnel JJs in batch A, respectively. Details on the experimental setup and procedures are reported in Sec. 1 in Supplementary Material. By linear fitting of the superconducting branch, we estimate the resistance $R_0$ as a function of the temperature $T$ (see dashed black lines in panels a) and b), as an example), plotted in panel c). The qualitative decrease of $R_0$ when decreasing $T$ is consistent with the analytical expression for $R_0(T)$,
\begin{equation}
\label{R0}
    R_0=2\pi/\Gamma(T) R_{env}/Qe^{-\Gamma(T)},
\end{equation}
where $R_{env}$ is the environment resistance~\cite{Kautz1990}. 
A direct transition from classical to quantum phase-diffusion, which manifests with a clear saturation of $R_0$ to some finite value, is expected to occur in the specific case of JJs characterized by $E_J/E_c$ ratios of the order of the unity, like in ultra-small JJs~\cite{Iansiti1987}, far below the energy regime of the devices here analyzed. 

Additional evidence of phase-diffusion processes in Al-based non-magnetic and magnetic JJs is provided by SCDs experiments, reported in Fig.~\ref{fig2}. Details on the experimental technique are reported in Sec. 1 in Supplementary Material.  In panels a) and b) in Fig.~\ref{fig2}, we report the Probability Density Distributions for each acquired temperature, for the SIS and SIsFS JJs of Batch B, respectively. The decreasing experimental behavior of $\sigma(T)$ is evidence of PD processes~\cite{Iansiti1987, Kautz1990, Vion1996, kivioja2005, mannik2005,krasnov2005,fenton2008,Yu2011,Yoon2011,longobardi2011}. This is further supported by the SCD skewness $\gamma$, also reported in panel c), which is consistent with symmetric SCD Probability densities~\cite{Iansiti1987, Kautz1990, Vion1996, kivioja2005, mannik2005,krasnov2005,fenton2008,Yu2011,Yoon2011,longobardi2011,Massarotti2015}. Independently of the presence of magnetic or non-magnetic barriers, both the superconducting branch resistance $R_0$ due to classical phase-diffusion and the standard deviation of SCDs follow not only the same thermal behavior but most importantly they are quantitatively consistent. As a matter of fact, the ratio between the standard deviation and the mean switching current $\sigma/I_{sw}$ at base temperature for both the magnetic and non-magnetic JJs is of the order of $4\%$, and does not depend on the nature of the barrier. Moreover, the simultaneous occurrence of PD effects both in the SCDs and the I-V characteristics is a quite rare phenomenon, occurring when $E_J$ is comparable to the thermal energy. This can be explained by assuming a frequency-dependent damping parameter~\cite{Iansiti1987,Kautz1990,mannik2005,kivioja2005,longobardi2012}, thus at frequencies comparable to the junction plasma frequency, the intrinsic resistance of the device (the subgap resistance $R_{sg}$ or the normal resistance $R_N$) plays no significant role if compared with the resistance of the environment ($R_{env}$) in which the JJs are embedded, since $R_{sg},R_N\gg R_{env}$~\cite{Kautz1990, Stornaiuolo2013}. This confirms the high quality of SIsFS also in extreme regimes of very low $E_J$ and does not pose any limitation to the integration of SIsFS JJs in quantum architectures. 

Taking as a reference the SIS JJ in Batch B characterized by the largest $I_c$ value, from the $(Q,\Gamma)$ phase diagram in Ref.~\cite{longobardi2012} a first estimation of the device high-frequency quality factor $Q$ can be derived. For $\Gamma=0.003$ at base temperature, the JJ does not show any clear transition to the MQT regime, nor the TA-like behavior of $\sigma$ at higher temperatures. These features are consistent with a quality factor of the order of $Q\sim1.3$ for this device, corresponding to the case where in the whole temperature range only PD effects have been observed, and MQT phenomena may arise only below the base temperature. Indeed, following the expression for the crossover temperature transition $T_{cr}$ in Eq.~\ref{Tcr}, $T_{cr}$ is of the order of $\sim2$ mK, consistently with what expected for low-$E_J$ SIS JJs. $T_{cr}$ has been derived by using the geometrical capacitance of the device, as well as the critical current in the absence of thermal fluctuations $I_{c0}=130$ nA, estimated by fitting the switching current probability density at base temperature, as detailed in Sec. 2 in Supplementary Material. This analysis allows us to determine the resistance of the environment from Eq.~\ref{Q}, which is of the order of $~10$ \text{$\Omega$} for the junctions reported in this work. This value is consistent with several studies in the PD regime~\cite{Devoret1985,Clarke1988,Kautz1990,kivioja2005,mannik2005,krasnov2005,longobardi2011,longobardi2012,Stornaiuolo2013} and further confirms that $R_{env}$ is much lower than the subgap resistance and the normal resistance of the tunnel JJs. 
Finally, considering that the electronics and cryogenic setup used for the SIsFS investigation are the same, we have used the value of $R_{env}$ to estimate self-consistently the quality factor of the magnetic tunnel SIsFS JJ. For a fitted  $I_{c0}=98$ nA for this JJ, $Q\sim1.1$. Similarly, we provide a rough estimation of the quality factors for JJs in batch A using the same approach, and we obtain $Q\sim0.4$ for the SIS JJ and $Q\sim0.35$ for the SIsFS JJ, respectively. For batch A, the expected $T_{cr}$ is of the order of $1$ mK. 

Since the SIsFS devices can be considered as a series between SIs and sFS JJs~\cite{vettoliere2022,Vettoliere2022aluminum}, our findings support the hypothesis that the electrodynamics of tunnel SIsFS JJs is mainly governed by the SIs tunnel JJ, with the critical current of the SIs junction much lower than the sFS part. The comparative analysis of the phase-dynamics in magnetic and non-magnetic JJs here reported demonstrates that the introduction of a ferromagnetic barrier in the system does not provide any sensitive impact on the dissipation mechanisms of the device. This important result adds to a previous comparative analysis of the temperature dependance of the superconducting energy gap in magnetic and non-magnetic JJs in Refs.~\cite{vettoliere2022,Vettoliere2022aluminum}, which demonstrated that also the superconducting gap of magnetic and non-magnetic devices are quantitatively consistent. As a matter of fact, the Al interlayer (s layer in the SIsFS structure) does not experience any exchange field, since a thin natural Al\text{O${}_x$} barrier decouples the Al from the ferromagnetic layer~\cite{Hao1990,Miao2014}, thus preventing typical proximity-induced reduction of the superconducting gap at the s/F interface in hybrid ferromagnetic JJs~\cite{Buzdin2005}, which may facilitate detrimental quasiparticles excitations in presence of microwave drives, i.\,e. in the standard regime in which quantum circuits are operated. The strong quantitative and qualitative agreement between the transport and dissipation mechanisms observed so far in magnetic and non-magnetic JJs indicates that dissipation mechanisms are not influenced by the presence of a ferromagnetic barrier in the device. For a theoretical estimation of the expected quasiparticles-induced relaxation times in a prototypal hybrid ferrotransmon including the JJs analyzed in this work, we refer to Ref.~\cite{Ahmad2023}. Although an experimental investigation of the phase-dynamics in presence of microwave drives, which is outside the main goal of this work, may add further insights into the dissipation mechanisms in hybrid ferromagnetic Josephson devices, the study reported here support the possibility to integrate unconventional tunnel SIsFS JJs in quantum superconducting architectures. 

\section{Supplementary Material}

The Supplementary Material includes information on the experimental cryogenic and room-temperature setup for I-V and SCDs experiments in this manuscript, and details on experimental methods, analysis and fitting of the SCD in the thermal limit.

\begin{acknowledgments}
The work was supported by the Pathfinder EIC 2023 project "FERROMON-Ferrotransmons and Ferrogatemons for Scalable Superconducting Quantum Computers"; “SQUAD—On-chip control and advanced read-out for superconducting qubit arrays” Programma STAR PLUS 2020, Finanziamento della Ricerca di Ateneo, University of Napoli Federico II; "SuperLink- Superconducting quantum-classical linked computing systems", call QuantERA2 ERANET COFUND, CUP B53C22003320005, the PNRR MUR project PE0000023-NQSTI and the PNRR MUR project CN 00000013-ICSC. H.G.A., D.M. (Davide Massarotti) and F.T. thank SUPERQUMAP project (COST Action CA21144).
\end{acknowledgments}

\section*{Data Availability Statement}

The data that support the findings of this study are available from the corresponding author upon reasonable request.

\section{Conflict of Interest statement}

The authors have no conflicts to disclose.

\end{document}